\documentclass[prl,twocolumn,showpacs,amsmath,amssymb,superscriptaddress]{revtex4}
\usepackage{pifont,amssymb,epsf,graphicx}
\begin{document}

\title{Domain Walls in Normal and Superconducting States of Iron Pnictides}

\author{Huaixiang Huang}

\affiliation{Texas Center for Superconductivity and Department of
Physics, University of Houston, Houston, Texas 77204, USA}
\affiliation{Department of Physics, Shanghai University , Shanghai
200444, China}

\author{Degang Zhang}
\affiliation{Texas Center for Superconductivity and Department of
Physics, University of Houston, Houston, Texas 77204, USA}

\author{Tao Zhou}
\affiliation{College of Science, Nanjing University of Aeronautics
and Astronautics, Nanjing 210016, China}
\affiliation{Texas Center
for Superconductivity and Department of Physics, University of
Houston, Houston, Texas 77204, USA}

\author{C. S. Ting}
\affiliation{Texas Center for Superconductivity and Department
of Physics, University of Houston, Houston, Texas 77204, USA}

\begin{abstract}

The electronic and magnetic structures in the normal and
superconducting states of iron pnictides are investigated by solving
self-consistently the Bogoliubov-de Gennes equation. It is shown
that strong electron correlations can induce  domain walls, which
separate  regions with different spin density wave orders. At zero
or low electron doping, $90^\circ$ domain walls are formed while
anti-phase domain walls are produced at higher electron doping. On
the domain walls, larger electron densities are always present. The
results agree qualitatively with recent observations of scanning
tunneling microscopy and superconducting quantum interference device
microscopy.

\end{abstract}

\pacs{74.70.Xa, 75.60.Ch, 74.25.-q, 74.81.-g}

\maketitle

The discovery of a new family of layered superconductors, i.e., the
FeAs-based superconductors, could offer a new avenue to explore the
mechanism of high temperature superconductivity [1-5]. Similar to
the cuprates, the parent compounds of the FeAs-based superconductors
also possess the antiferromagntic ground state [4,5]. With
increasing electron or hole doping, antiferromagntic order is
suppressed and superconductivity appears in both the cuprates and
the iron pnictides. However, different from the cuprates, due to the
the strong nesting effect between the hole Fermi surfaces around the
$\Gamma$ point and the electron Fermi surfaces around the $M$ point
, superconductivity and the spin density wave (SDW) orders can
coexist in the electron-doped FeAs-based superconductors [6,7].
Because each unit cell of the FeAs-based superconductors contains
two inequivalent Fe ions, different arrays of magnetic moments on Fe
ions in both normal and superconducting states could lead to diverse
magnetic structures and uncommon electronic properties [8,9].

Recently, in scanning tunneling microscopy (STM) experiments twin
boundaries were observed in the normal state of
$\mathrm{Ca(Fe_{1-x}Co_x)_2As_2}$ [10]. Across these twin
boundaries, the $a$ ($b$) axis of the crystal rotates through
$90^\circ$. This means that the modulation direction of SDW is
rotated by $90^\circ$. In other words, $90^\circ$ domain walls are
formed at the twin boundaries. In Ref. [11], superconducting quantum
interference device microscopy (SQIDM) revealed that in the
superconducting state of underdoped
$\mathrm{Ba(Fe_{1-x}Co_x)_2As_2}$ with $x<0.07$, the diamagnetic
susceptibility is increased and the superfluid density is enhanced
on the twin boundaries or $90^\circ$ domain walls. In another STM
experiment [12], Li {\it et al.} also observed a $90^\circ$
anti-phase domain wall in the parent compounds of iron pnictides, on
which the local density of states (LDOS) is much higher than that in
the interior of magnetic domains. Therefore, domain walls exist
universally in underdoped FeAs-based superconductors and affect
strongly the electronic properties in the normal and superconducting
states.

In this work, we study the complex electronic and magnetic
structures in the underdoped FeAs-based superconductors by solving
self-consistently the Bogoliubov-de Gennes (BdG) equation. We start
from the model Hamiltonian $H=H_{0}+H_{SC}+H_{int}$. Here $H_{0}$ is
the two-orbital four-band tight-binding model proposed in Ref. [13],
which describes correctly the characteristics of the energy band
structure for the FeAs-based superconductors [14-19]. The hopping
parameters $t_1$-$t_4$ in $H_0$ are depicted in Fig. 1.

\begin{figure}
\rotatebox[origin=c]{0}{\includegraphics[angle=0,
           height=2in]{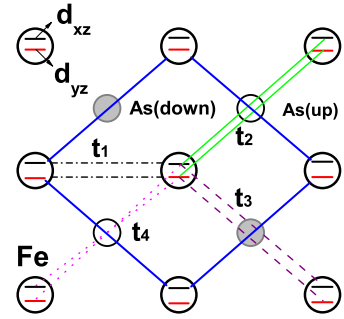}}
\caption {(Color online) Schematic of the tight-binding model $H_0$
in Ref. [13]. $t_1$ is the nearest neighbor hopping between the same
orbitals $d_{xz}$ or $d_{yz}$ on Fe ions, $t_2$ and $t_3$ are the
next-nearest neighbor hoppings between the same orbitals mediated by
the up and down As ions, respectively, and $t_4$ is the next nearest
neighboring hopping between the different orbitals.}

\end{figure}

The pairing Hamiltonian
$$H_{SC}=\sum_{{\bf i}\mu {\bf j}\nu}(\Delta_{{\bf i}\mu {\bf
j}\nu}c^\dagger_{{\bf i}\mu\uparrow}c^{\dagger}_{{\bf
j}\nu\downarrow}+{\rm h.c.}), \eqno{(1)}$$ where $\Delta_{{\bf i}\mu
{\bf j}\nu}$ is the pairing between the orbital $\mu$ ($d_{xz}$ or
$d_{yz}$) on the site ${\bf i}$ and the orbital $\nu$ ($d_{xz}$ or
$d_{yz}$) on the site ${\bf j}$, and $c^\dagger_{{\bf i}\mu\sigma}$
is the creation operator of an electron with spin $\sigma$ at the
orbital $\mu$ on the site ${\bf i}$.

The interaction Hamiltonian $H_{int}$ considered here only includes
on-site Coulomb interaction $U$ and Hund coupling $J_{H}$. After
taking the mean field treatment, $H_{int}$ can be expressed as [20]
$$H_{int}=U\sum_{{\bf i},\mu,\sigma\neq\bar{\sigma}}\langle
n_{{\bf i}\mu\bar{\sigma}}\rangle n_{{\bf
i}\mu\sigma}+(U-3J_H)\sum_{{\bf i},\mu\neq\nu,\sigma} \langle
n_{{\bf i}\mu\sigma}\rangle n_{{\bf i}\nu\sigma}$$
$$+(U-2J_H)\sum_{{\bf
i},\mu\neq\nu,\sigma\neq\bar{\sigma}} \langle n_{{\bf
i}\mu\bar{\sigma}}\rangle n_{{\bf i}\nu{\sigma}}, \eqno{(2)}$$
 where $n_{{\bf i}\mu\sigma}=c^\dagger_{{\bf i}\mu\sigma}c_{{\bf
i}\mu\sigma}$ .

We note that based on the model Hamiltonian $H$, the obtained
LDOS[13,20], phase diagram [20], and spin susceptibility at
different doping and temperature [21] for electron doped FeAs-based
superconductors, and Andreev bound states at negative energy inside
the vortex core [22] for hole doped FeAs-based superconductors are
all consistent with the STM [23-26], nuclear magnetic resonance
[6,7], and neutron scattering experiments [27-30].

The eigenvalues and eigenfunctions of $H$ can be obtained by solving
self-consistently the BdG equation, i.e.

$$\sum_{{\bf j},\nu} \left( \begin{array}{cc}
 H_{{\bf i}\mu {\bf j}\nu\sigma} & \Delta_{{\bf i}\mu {\bf j}\nu}  \\
 \Delta^{*}_{{\bf i}\mu {\bf j}\nu} & -H^{*}_{{\bf i}\mu {\bf j}\nu\bar{\sigma}}
\end{array}
\right) \left( \begin{array}{c} u^{n}_{{\bf
j}\nu\sigma}\\v^{n}_{{\bf j}\nu\bar{\sigma}}
\end{array}
\right) =E_n \left( \begin{array}{c} u^{n}_{{\bf
i}\mu\sigma}\\v^{n}_{{\bf i}\mu\bar{\sigma}}
\end{array}
\right),\eqno{(3)}$$
 where $H_{{\bf i}\mu {\bf j}\nu\sigma}$ is the matrix element of
 $H$ with spin $\sigma$ between the orbital $\mu$ on the site ${\bf
 i}$ and the orbital $\nu$ on the site ${\bf j}$. The
 superconducting pairing $\Delta_{{\bf i}\mu {\bf j}\nu}\equiv \frac{1}{2}\langle
c_{\bf{i}\mu\uparrow}c_{\bf{j}\nu\downarrow}-c_{\bf{i}\mu\downarrow}c_{\bf{j}\nu\uparrow}\rangle$
in real space is associated with the eigenvalues $E_n$ and the
 eigenfunctions $(u^{n}_{{\bf i}\mu\sigma},v^{n}_{{\bf
 i}\mu\bar{\sigma}})$, and has the form

$$\Delta_{{\bf i}\mu {\bf j}\nu}=\frac{V_{{\bf i}\mu {\bf
j}\nu}}{4}\sum_n (u^{n}_{{\bf i}\mu\uparrow}v^{n*}_{{\bf
j}\nu\downarrow}+u^{n}_{{\bf j}\nu\uparrow}v^{n*}_{{\bf
i}\mu\downarrow})\tanh (\frac{E_n}{2k_B T}) \eqno{(4)}$$
 at  temperature $T$. Here, $k_B$ is the Boltzmann's constant,
and $V_{{\bf i}\mu {\bf j}\nu}$ is the pairing potential between the
orbitals on the sites ${\bf i}$ and ${\bf j}$. The corresponding
local electron density reads

$$ n_{{\bf i}} =\sum_{n,\mu} \{|u^{n}_{{\bf
i}\mu\uparrow}|^{2}f(E_n)+|v^{n}_{{\bf
i}\mu\downarrow}|^{2}[1-f(E_n)]\},\eqno{(5)}$$ where $f(E_n)$ is the
Fermi function, and the local magnetic moment
$m_i=\frac{1}{2}\sum_{\mu}(\langle n_{i\mu\uparrow}\rangle-\langle
n_{i\mu\downarrow}\rangle)$.

In order to interpret the complex domain wall structures seen by STM
experiments on the iron pnictides, we investigate the strong Coulomb
correlation on Fe sites. In our calculations, we have employed the
hopping parameters in Ref. [13], i.e. $t_1=1$, $t_2=0.4$,
$t_3=-2.0$, and $t_1=0.04$, and have chosen $U=4.8$, $J_H=1.3$, and
$V_{{\bf i}\mu {\bf j}\nu}=1.1$ for $\mu=\nu$ and $|{\bf i}-{\bf
j}|=\sqrt{2}$, and 0 for all other cases. Note that only the
electron pairings between the same orbitals on the next nearest
neighboring Fe sites are considered. Such a choice of the pairing
potential leads the superconducting order parameter to be
$s_\pm$-wave type.

\begin{figure}
\rotatebox[origin=c]{0}{\includegraphics[angle=0,
           height=4.1in]{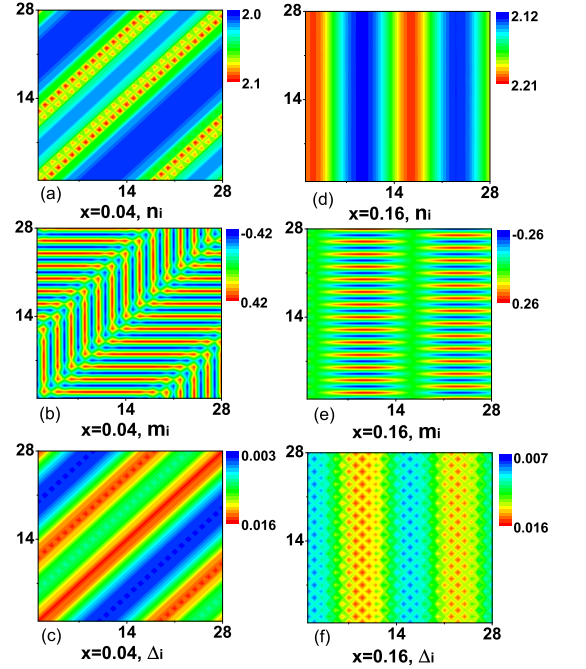}}
\caption {(Color online) The images of electron density $n_i$,
magnetic order $m_i$, and superconducting order $\Delta_i$ at
electron dopings $x=0.04$ and 0.16.}
\end{figure}

In Fig. 2, we present the zero temperature images of electron
density $n_i$, magnetic order $m_i$, and superconducting order
$\Delta_i\equiv\frac{1}{8}\sum_{{\bf j},\mu}\Delta_{{\bf i}\mu {\bf
j}\mu}$ at electron dopings $x=0.04$ and 0.16 on a $28\times 28$
lattice with period boundary condition. From Fig. 2 (b) and (e), we
can see that there exist obviously  magnetic domain structures.
Across the domain walls, the modulation direction of magnetic order
rotates through $90^\circ$ at $x=0.04$ while the phase of magnetic
order changes sign at $x=0.16$. So the anti-phase domain walls,
predicted previously in Ref. [8], are realized in the higher
electron doped case. We observe that on both $90^\circ$ domain walls
and anti-phase domain walls, there are always higher electron
densities $n_i$, as shown in Fig. 2 (a) and (d). Therefore, it is
expected that superfluid density is enhanced on these domain walls,
which coincides with the observations of SQIDM experiments [11].
However, the superconducting order parameter $\Delta_i$ has a larger
magnitude on $90^\circ$ domain walls, but has a smaller magnitude on
anti-phase domain walls (see. Fig.2 (c) and (f)).

In order to see clearly the variations of $n_i$, $m_i$, and
$\Delta_i$ with  distance from the domain walls, in Fig. 3, we give
their values on the line $y=14$. Fig. 3(a) and (d) show that more
electrons are accumulated near all the domain walls. On $90^\circ$
domain walls at $x=0.04$ in Fig. 3(b) and (c), both $m_i$ and
$\Delta_i$ have their maximum values. Oppositely, on anti-phase
domain walls at $x=0.16$ in Fig. 3(e) and (f), $m_i$ almost
vanishes, and $\Delta_i$ has the minimum values. Therefore, except
for the electron density, the magnetic and superconducting
properties on the two kinds of domain walls are very different.

We would like to mention that when $x<\sim 0.02$, $n_i$ and $m_i$
have similar patterns with those at $x=0.04$, except $\Delta_i=0$.
Therefore, the $90^\circ$ domain wall structure also exists in the
normal state of the FeAs-based superconductors, and is consistent
with the observations of STM experiments [10]. However, we do not
get the solution of complex $90^\circ$ anti-phase domain wall seen
in the parent compounds [12], which cannot be formed under the
period boundary condition.

\begin{figure}
\rotatebox[origin=c]{0}{\includegraphics[angle=0,
           height=3in]{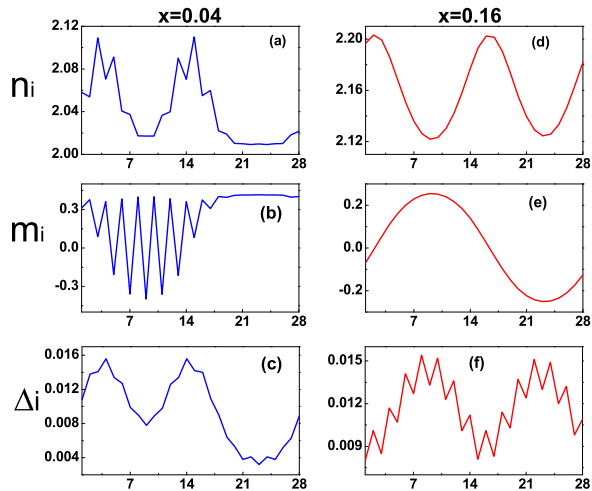}}
\caption{(Color online) Electron density $n_i$, magnetic order
$m_i$, and superconducting order $\Delta_i$ on the line $y=14$ at
electron dopings $x=0.04$ and 0.16.}
\end{figure}

Our calculations also show that with increasing electron doping, the
magnetic order $m_i$ is gradually suppressed and finally vanishes at
$x\sim 0.2$, which is larger than the experimental value. The main
difference between the theoretical results and experimental data
could be due to the fact that a strong Coulomb interaction $U$ leads
to renormalization of the hole Fermi surfaces around the $\Gamma$
point and the electron Fermi surfaces around the $M$ point in the
FeAs-based superconductors, which enhances the nesting effect
between the hole Fermi surfaces and the electron Fermi surfaces.
However, by adjusting suitably the hopping parameters $t_2$ and
$t_3$ in $H_0$, which determine the sizes and shapes of the Fermi
surfaces, the nesting effect could be diminished and the
experimental value can be obtained. When $x<\sim 0.08$, the
$90^\circ$ domain walls always exist while the anti-phase magnetic
domains show up at $x>\sim 0.15$. When $\sim 0.08<x<\sim 0.15$, SDW
and superconductivity uniformly coexist. We also note that the
solution of anti-phase domain wall structure is a metstable state in
the above range, which has a slightly higher energy than the ground
state.

Now we calculate the LDOS on and near the domain walls in order to
compare with the STM experiments. The expression of LDOS  at energy
$\omega$ on the site ${\bf i}$ is

$$ \rho_{\bf i}(\omega)=\frac{1}{N}\sum_{n\mu
{\bf k}}\{|u^{n}_{{\bf i}\mu\uparrow{\bf k}}|^{2}\delta(E_{n{\bf
k}}-\omega) +|v^{n}_{{\bf i}\mu\downarrow{\bf
k}}|^{2}\delta(E_{n{\bf k}}+\omega)\}, \eqno{(6)}$$
 where $N$ is the number of wave vectors ${\bf k}$, $(u^{n}_{{\bf i}
\mu\uparrow{\bf k}},v^{n}_{{\bf i}\mu\downarrow{\bf k}})$ and
$E_{n{\bf k}}$ are the eigenfunctions and eigenvalues of the Fourier
transformed BdG equation, respectively. In calculating $\rho_{\bf
i}(\omega)$, we have taken the delta function
$\delta(x)=\Gamma/\pi(x^2+\Gamma^2)$ with quasipartical damping
$\Gamma=0.005$, and a $30\times 30$ supercell is used.

\begin{figure}
\rotatebox[origin=c]{0}{\includegraphics[angle=0,
           height=3.2in]{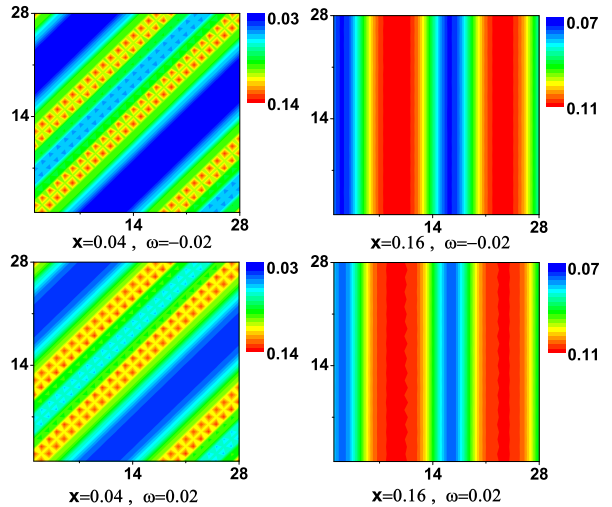}}
\caption{(Color online) The images of LDOS $\rho_{\bf i}(\omega)$
with electron dopings $x=0.04$ and 0.16 at energies $\omega=\pm
0.02$.}
\end{figure}

Fig. 4 shows the LDOS images on a $28\times 28$ lattice at different
energies and electron dopings. It is obvious that when $\omega=\pm
0.02$, $\rho_{\bf i}(\omega)$ has the maximum value on the
$90^\circ$ domain walls at $x=0.04$, but has the minimum value on
the ant-phase domain walls at $x=0.16$. Here we also omit the LDOS
image with $x=0.0$, which is similar to that of $x=0.04$.

However, the LDOS images change with energy $\omega$. In Fig. 5, we
give the energy dependence of the LDOS $\rho_{\bf i}(\omega)$ at the
sites on and near domain walls with different electron dopings. In
both $x=0.0$ and 0.04, obviously, $\rho_{\bf i}(\omega)$ at
$(14,14)$ on the $90^\circ$ domain wall is always larger than that
on the other sites when $\omega\in (-0.18,0.1)$. In contrast,
$\rho_{\bf i}(\omega)$ at $(1,14)$ on the anti-phase domain wall is
always smaller than that on the other sites when $\omega > -0.13$.
We note that when $\omega\in (-0.2,0.2)$, the curves of $\rho_{\bf
i}(\omega)$ with $x=0.0$ resemble those measured by the STM
experiments in the parent compounds [25]. The coherence peak at
positive energy is higher at both $x=0.04$ and 0.16 due to the
coexistence of SDW and superconductivity [13,20]. The asymmetry of
the coherence peaks was also observed by the STM experiments
[23-25].

\begin{figure}
\rotatebox[origin=c]{0}{\includegraphics[angle=0,
           height=1.8in]{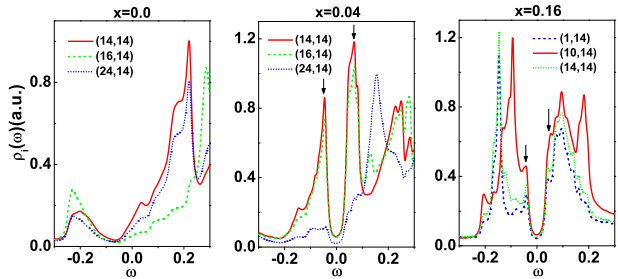}}
\caption{(Color online) Energy dependence of LDOS $\rho_{\bf
i}(\omega)$ with electron dopings $x=0.0$, 0.04, and 0.16 at the
sites on and near domain walls. The arrows point to the coherence
peaks with maximum superconducting order parameter $\Delta_i$ in the
panels of $x=0.04$ and 0.16 }
\end{figure}

In summary, we have studied the  electronic and magnetic properties
in the electron underdoped iron pnictides. Due to strong electron
correlations, the domain walls are formed at the twin boundaries
producing below the structural transition of the parent compounds.
The existence of the domain wall structures leads to the
nonuniformity of the electron density and the superconducting order
parameter in real space. Therefore, inhomogeneity of
superconductivity is intrinsic in the underdoped iron pnictides. The
$90^\circ$ domain walls and the supercurrent properties on them have
been confirmed by the STM and SQIDM experiments. However, the
anti-phase domain walls at the higher electron doping are not
experimentally reported yet. We hope that such a magnetic structure
could be verified by future STM experiments.

The authors would like to thank S. H. Pan, Ang Li and Yi Gao for
useful discussions. This work was supported by the Texas Center for
Superconductivity at the University of Houston and by the Robert A.
Welch Foundation under Grant No. E-1146. Huang also acknowledge the
support of Shanghai Leading Academic Discipline Project (project
number S30105) and Shanghai Education Development Project.


\begin{thebibliography}{99}


\bibitem{1} Y. Kamihara, T. Watanabe, M. Hirano and H. Hosono, J. Am. Chem. Soc. {\bf 130}, 3296 (2008).
\bibitem{2} Zhi-An Ren, Wei Lu, Jie Yang, Wei Yi, Xiao-Li Shen, Zheng-Cai Li, Guang-Can Che, Xiao-Li Dong,
Li-Ling Sun, Fang Zhou,and Zhong-Xian Zhao, Chin. Phys. Lett. {\bf
25}, 2215 (2008).
\bibitem{3} X. H. Chen, T. Wu, G. Wu, R. H. Liu, H. Chen, and D. F. Fang , Nature (London) {\bf 453}, 761 (2008).
\bibitem{4} C. de la Cruz, Q. Huang, J. W. Lynn, Jiying Li, W. Ratcliff II, J. L. Zarestky, H. A. Mook,
G. F. Chen, J. L. Luo, N. L. Wang, and Pengcheng Dai , Nature
(London) {\bf 453}, 899 (2008).
\bibitem{5} G. F. Chen, Z. Li, D. Wu, G. Li, W. Z. Hu, J. Dong, P. Zheng, J. L. Luo, and N. L. Wang,
Phys. Rev. Lett. {\bf 100}, 247002 (2008).
\bibitem{6} Y. Laplace, J. Bobroff, F. Rullier-Albenque, D. Colson, and A. Forget, Phys. Rev. B {\bf 80}, 140501 (2009).
\bibitem{7} M.-H. Julien, H. Mayaffre, M. Horvatic, C. Berthier, X. D. Zhang, W. Wu, G.F. Chen, N.L. Wang,
and J.L. Luo , Europhys. Lett. {\bf 87}, 37001 (2009).
\bibitem{8} I. I. Mazin and M. D. Johannes, Nature Phys. {\bf 5}, 141 (2009).
\bibitem{9} Lev P. Gor'kov and Gregory B. Teitel'baum, arXiv:1001.4641.
\bibitem{10} T.-M. Chuang, M. P. Allan, Jinho Lee, Yang Xie, Ni Ni, S. L. Bud¡¯ko,
G. S. Boebinger, P. C. Canfield, and J. C. Davis,  Science {\bf
327}, 181 (2010).
\bibitem{11} B. Kalisky, J. R. Kirtley, J. G. Analytis, J.-H Chu, A. Vailionis, I. R. Fisher,
and K. A. Moler , Phys. Rev. B {\bf 81}, 184513 (2010).
\bibitem{12} Guorong Li, Xiaobo He, Ang Li, Shuheng H. Pan, Jiandi Zhang, Rongying Jin, A. S. Sefat, M. A. McGuire,
D. G. Mandrus, B. C. Sales, and E. W. Plummer, arXiv:1006.5907.
\bibitem{13} Degang Zhang, Phy. Rev. Lett. {\bf 103}, 186402 (2009); {\it ibid},
{\bf 104}, 089702 (2010).
\bibitem{14} H. Ding, P. Richard, K. Nakayama, T. Sugawara, T. Arakane, Y. Sekiba, A. Takayama, S. Souma,
T. Sato, T. Takahashi, Z. Wang, X. Dai, Z. Fang, G. F. Chen, J. L.
Luo, and N. L. Wang, Europhys. Lett. {\bf 83}, 47001 (2008).
\bibitem{15} D. H. Lu, M. Yi, S.-K. Mo, A. S. Erickson, J. Analytis, J.-H. Chu, D. J. Singh, Z. Hussain,
T. H. Geballe, I. R. Fisher, and Z.-X. Shen, Nature (London) {\bf
455}, 81 (2008).
\bibitem{16} C. Liu, G. D. Samolyuk, Y. Lee, N. Ni, T. Kondo, A. F. Santander-Syro, S. L. Bud'ko,
J. L. McChesney, E. Rotenberg, T. Valla, A. V. Fedorov, P. C.
Canfield, B. N. Harmon, and A. Kaminski , Phys. Rev. Lett. {\bf
101}, 177005 (2008).
\bibitem{17} T. Kondo, A. F. Santander-Syro, O. Copie, Chang Liu, M. E. Tillman, E. D. Mun,
J. Schmalian, S. L. Bud'ko, M. A. Tanatar, P. C. Canfield, and A.
Kaminski , Phys. Rev. Lett. {\bf 101}, 147003 (2008).
\bibitem{18} V. B. Zabolotnyy, D. S. Inosov, D. V. Evtushinsky, A. Koitzsch, A. A. Kordyuk, G. L. Sun,
J. T. Park, D. Haug, V. Hinkov, A. V. Boris, C. T. Lin, M. Knupfer,
A. N. Yaresko, B. Buechner, A. Varykhalov, R. Follath, and S. V.
Borisenko , Nature (London) {\bf 457}, 569 (2009).
\bibitem{19} K. Terashima, Y. Sekiba, J. H. Bowen, K. Nakayama, T. Kawahara, T. Sato, P. Richard,
Y.-M. Xu, L. J. Li, G. H. Cao, Z.-A. Xu, H. Ding, and T. Takahashi,
PNAS {\bf 106}, 7330 (2009) .
\bibitem{20} Tao Zhou, Degang Zhang, and C. S. Ting, Phys. Rev. B {\bf 81}, 052506 (2010).
\bibitem{21} Yi Gao, Tao Zhou, C. S. Ting, and Wu-Pei Su, arXiv:1003.2609.
\bibitem{22} Yi Gao {\it et al.}, in preparation.
\bibitem{23} Y. Yin, M. Zech, T. L. Williams, X. F. Wang, G. Wu, X. H. Chen,
and J. E. Hoffman, Phys. Rev. Lett. {\bf 102}, 097002 (2009).
\bibitem{24} F. Massee, Y. Huang, R. Huisman, S. de Jong, J.B. Goedkoop,
and M.S. Golden, Phys. Rev. B {\bf 79}, 220517(R) (2009).
\bibitem{25} Ang Li {\it et al.}, (to be published).
\bibitem{26} Lei Shan, Yong-Lei Wang, Bing Shen, Bin Zeng, Yan Huang, Ang Li, Da Wang, Huan Yang,
Cong Ren, Qiang-Hua Wang, Shuheng Pan, and Hai-Hu Wen ,
arXiv:1005.4038.
\bibitem{27} M. D. Lumsden, A. D. Christianson, D. Parshall, M. B. Stone, S. E. Nagler, G.J. MacDougall,
H. A. Mook, K. Lokshin, T. Egami, D. L. Abernathy, E. A.
Goremychkin, R. Osborn, M. A. McGuire, A. S. Sefat, R. Jin, B. C.
Sales, and D. Mandrus, Phys. Rev. Lett. {\bf 102}, 107005 (2009).
\bibitem{28} D. K. Pratt, W. Tian, A. Kreyssig, J. L. Zarestky, S. Nandi, N. Ni, S. L. Bud¡¯ko,
P. C. Canfield, A. I. Goldman, and R. J. McQueeney
, Phys. Rev. Lett. {\bf 103}, 087001 (2009).
\bibitem{29} A. D. Christianson, M. D. Lumsden, S. E. Nagler, G. J. MacDougall, M. A. McGuire,
A. S. Sefat, R. Jin, B. C. Sales, and D. Mandrus, Phys. Rev. Lett.
{\bf 103}, 087002 (2009).
\bibitem{30} D. S. Inosov, J. T. Park, P. Bourges, D. L. Sun, Y. Sidis, A. Schneidewind,
K. Hradil, D. Haug, C. T. Lin, B. Keimer, and V. Hinkov, Nature
Phys. {\bf 6}, 178 (2010).


\end{thebibliography}
\end{document}